# Malware in the future? forecasting of analyst detection of cyber events

Jonathan Z. Bakdash[1,2*], Steve Hutchinson[3], Erin G. Zaroukian[4], Laura R. Marusich[5], Saravanan Thirumuruganathan[6], Charmaine Sample[3], Blaine Hoffman[4], and Gautam Das[7]

[1]U.S. Army Research Laboratory South at the
University of Texas Dallas
Richardson, TX, USA

[2]Department of Psychology, Counseling, and Special Education
Texas A&M Commerce
Commerce, TX, USA

[3]Computational and Information Sciences Directorate
ICF for the U.S. Army Research Laboratory
Adelphi, MD, USA

[4]Human Research and Engineering Directorate
U.S. Army Research Laboratory
Aberdeen Proving Ground, MD, USA

[5]U.S. Army Research Laboratory South at the
University of Texas Arlington
Arlington, TX, USA

[6]Qatar Computing and Research Institute
Qatar Foundation
Doha, Qatar

[7]Computer Science and Engineering Department
University of Texas Arlington
Arlington, TX, USA

***Corresponding author**: E-mail: jonathan.z.bakdash.civ@mail.mil

**Key words:** cybersecurity; forecasting; prediction; cyber attack; malware; computer security service provider

# Abstract

Cyber attacks endanger physical, economic, social, and political security. There have been extensive efforts in government, academia, and industry to anticipate, forecast, and mitigate such cyber attacks. A common approach is time-series forecasting of cyber attacks based on data from network telescopes, honeypots, and automated intrusion detection/prevention systems. This research has uncovered key insights such as systematicity in cyber attacks. Here, we propose an alternate perspective of this problem by performing forecasting of attacks that are *analyst-detected* and *-verified* occurrences of malware. We call these instances of malware cyber event data. Specifically, our dataset was analyst-detected incidents from a large operational Computer Security Service Provider (CSSP) for the U.S. Department of Defense, which rarely relies only on automated systems. Our data set consists of weekly counts of cyber events over approximately seven years. This curated dataset has characteristics that distinguish it from most datasets used in prior research on cyber attacks. Since all cyber events were validated by analysts, our dataset is unlikely to have false positives which are often endemic in other sources of data. Further, the higher-quality data could be used for a number of important tasks for CSSPs such as resource allocation, estimation of security resources, and the development of effective risk-management strategies. To quantify bursts, we used a Markov model of state transitions. For forecasting, we used a Bayesian State Space Model and found that events one week ahead could be predicted with reasonable accuracy, with the exception of bursts. Our findings of systematicity in analyst-detected cyber attacks are consistent with previous work using cyber attack data from other sources. The advanced information provided by a forecast may help with threat awareness by providing a probable value and range for future cyber events one week ahead, similar to a weather forecast. Other potential applications for cyber event forecasting include proactive allocation of resources and capabilities for cyber defense (e.g., analyst staffing and sensor configuration) in CSSPs. Enhanced threat awareness may improve cybersecurity by helping to optimize human and technical capabilities for cyber defense.

# Introduction

Internet infrastructure plays a crucial role in a number of daily activities. The pervasive nature of cyber systems ensures far-reaching consequences of cyber attacks. Cyber attacks threaten physical, economic, social, and political security. The effects of cyber attacks can disrupt, deny, and even disable the operation of critical infrastructure including power grids, communication networks, hospitals, financial institutions, and defense and military systems. To protect its critical infrastructure, the U.S. Department of Defense (DoD) has identified cyberspace (information networks for computers, communication, and other systems) as a key operational environment for the military, one that is interdependent with the physical (air, land, maritime, and space) environment [1]. A key component of the DoD's strategy and implementation plans for protecting cyberspace is enhancing threat awareness for Computer Security Service Providers[1] [CSSPs] [2,3]. Analysts in DoD CSSPs protect DoD and DoD affiliated computers and networks by finding, analyzing, remediating, and documenting cyber attacks.

To improve threat awareness for CSSPs, we investigate whether intrinsic, predictable patterns exist among *analyst-detected* and *-verified* occurrences of malware, referred to here as cyber events. This research is unique because the dataset comprises over seven years of cyber events from an operational DoD CSSP that rarely relies only on automated systems (e.g., anti-virus software, firewalls, intrusion detection systems, and intrusion protection systems which are typically signature based) to detect and verify attacks.

In contrast, nearly all prior research on modeling cyber attacks [4–10] lacked analyst detection and verification of computer security incidents with the exceptions of [11,12]. In these two exceptions, security incidents were verified by system administrators at a large university [11] or verified by analysts at a CSSP [12]. Thus in most earlier research, the sources for cyber attacks were processed data from network telescopes and honeypots [4,6,8–10] and alerts from automated systems on real networks [5,7,13]. Compared to real networks, the majority of traffic to network telescopes (passive monitoring of unrequested network traffic to unused IP addresses) and honeypots (monitored and isolated systems that are designed to appear legitimate to attackers) can be considered suspicious. However, the rates of false alarms and misses for cyber attacks depend upon how "attacks" are determined in data processing and/or the automated systems [14]. Cyber attacks inferred from automated systems generate a large volume of alerts and are considered a standard measure for attacks [15]. Yet automated systems only indicate an attack *may* have occurred because of false positives and misses [12,16]. Many automated systems compare information gathered by sensors to patterns stored as signatures, looking for matches. A signature that is too specific may miss attacks, just as one that is not specific enough may generate false positives. Consequently, there are multiple approaches to manage the volume of alerts by filtering and combining them (e.g., flows of related alerts) [7,13]. Other work on improving IDS detection has applied machine learning [17,18] to a widely used synthetic dataset on cyber attacks, KDDCUP 1999 [19]. The original dataset contained a high number of duplicate records, which inflated classification accuracy and was addressed in a new version of the dataset, NSL-KDD [18]. Nonetheless, the KDDCUP 1999 dataset and derivatives of it are now quite old and are likely not representative of current systems.

Despite potential limitations on the data sources in the majority of prior research, intrinsic patterns in cyber attacks have been identified. The main finding is that cyber attack frequency can be forecast over time using processed data from network telescopes and honeypots as well as automated systems [4,6,7–10,20]. Similarly, the only research using human-verified cyber attack data also found intrinsic temporal patterns [11,12]. Other patterns have also been found, including the presence of bursts or extreme values [6,9,10] and disproportionate exploitation of specific vulnerabilities [5]. Taken together, this work suggests that cyber attacks have a deterministic component: They are not fully stochastic (or random point processes).

---

[1] Note that CSSPs are also be referred to as Computer Defense Service Providers, Computer Network Defense Service Providers, Computer Security Service Providers, Cybersecurity Defense, Managed Computer Security Service Providers, and Managed Security Service Providers. The last two terms, which included "managed," explicitly refer to a CSSP that protects the networks and computers for multiple clients/customers.

The current work is novel because little is known about the systematicity for analyst detected and verified cyber attacks protecting critical infrastructure, in this case, U.S. military networks, including organizations affiliated with the U.S. Department of Defense. The curated analyst dataset is unlikely to suffer from false positives because cyber events are detected and verified by analysts who investigate events and connect evidence to confirm or disconfirm potential events, also see [12]. Consequently, analyst-detected and -verified cyber events provide a potentially more direct, filtered, and informative indicator of threats for CSSPs than attack data processed from network telescopes and honeypots and alerts from automated systems in isolation. However, in our data and other real-world datasets, whether with automated systems or analysts, the number of attacks that were missed is unknown. A miss could be an attack that was completely overlooked or delayed detection.

In this paper, we make three crucial contributions. First, as previously described, we advocate for the forecasting of cyber incidents from analysts, as it has a number of appealing properties over purely machine-processed data using rules. Second, we describe characteristics of the data: its overdispersion and extreme values or bursts in cyber event counts. We quantify bursts using a Markov model. Bursts are a signature of natural phenomena, including human behavior [21]. Last, we perform temporal prediction of cyber events using the Bayesian State Space Model (BSSM) to predict the number of future events one week ahead. This approach provides both a point estimate and also an interval for the range of forecast uncertainty. Predictive models of analyst-based cyber events may proactively inform CSSPs on a number of important tasks such as resource allocation (e.g., number and location of sensors on the network), estimation of analyst staffing, and the development of effective risk-management strategies

## Related Research

Prior work has demonstrated that cyber attacks, predominantly using machine processed data or alerts, exhibit both deterministic and stochastic patterns. The main finding for deterministic patterns in cyber attacks is that they are neither independent nor random over time. Consequently, the number of attacks in the past helps predict the number of future attacks. The deterministic patterns can often be leveraged to generate reasonably accurate predictions. When stochastic patterns are present, particularly fluctuations such as extreme values and bursts, temporal forecasting becomes more challenging. The majority of time-series models for count data (i.e., number of attacks) assume a Poisson distribution (equal rate and variance, expressed in a single parameter) [22]. Analysis of automated system data of cyber attacks also reveals systematicity in the tendency of attacks to exploit a disproportionately small set of vulnerabilities.

### Forecasting and bursts

Time-series forecasting has been widely used for prediction of cyber attacks. Using machine-processed attack data from network telescopes and honeypots, the number of cyber attacks over time at minute and hour intervals are predictable over the time period of up to one day [6,8–10]. Also, cyber attacks were modeled at different levels (attacker IP address, targeted network port[s], and victim IP address) [10]. The use of different levels for attacks were extended to an early warning system by modeling multiple time-series for attack penetration and the number of victims [8]. Other research has reported temporal patterns in attack, but instead used filtered and combined alerts from automated systems [7,13] or used publically reported cyber attacks from the Hackmageddon database [23]. Last, extreme values—bursts—in attack frequencies have been identified and used to improve model prediction [6,9,10]. While extreme values in cyber attacks pose a challenge for accurate time-series forecasting, such bursts also underlie human behavior [21].

Bursts in cyber attacks, however, are not a universal pattern. Using analyst verified reports, bursts of cyber attacks were found in only three out of five customer computers/networks protected by a CSSP [12]. Similarly, bursts were not reported for distributed denial-of-service (DDoS) attacks, but these data were limited to one-minute intervals over less than an hour [24]. The mixed finding for the presence of bursts, or lack thereof, raises the question of determining conditions for their occurrence [12]. It is possible that the absence of bursts may be due to insufficient data, the specific method(s) of attack detection, and/or differences in the aggregation/source of the data (e.g., single versus multiple CSSP customers, the type organization for the computers and networks being protected). Nonetheless, when extreme values of attacks are present, and they are not completely random, including them in modeling improves prediction accuracy [6,9,10,12].

**Vulnerabilities**

Non-random exploitation of vulnerabilities provides converging evidence for systematicity in cyber attacks. A small number of vulnerabilities tend to comprise the majority of exploits. For example, data breaches can be classified with 90% accuracy using two types of externally observable risks: a) misconfigured internal systems (e.g., not changing the default username and password) and b) anomalous outbound traffic (e.g., spam, port scanning) [25]. Another approach used internal network monitoring logs to identify the probabilities of malware using specific vulnerable vectors (e.g., network configuration, unpatched software, and particular services) [5]. In addition to vulnerability, predictability can be seen in the small number of "attacker" IP addresses, or points of origin, which account for the majority of cyber attacks [4].

## Dataset description

The dataset here comprises 9,302 cyber event reports of malware. In general, malware includes computer viruses, worms, trojan horses, malicious mobile code, and blended attacks (combinations of the previous methods) [26]. Possible attack vectors for delivering malware include email attachments, web browsers (e.g., websites with malicious javascript code or other embedded malicious code), user installed software, and removable media (e.g., flash drives) [26]. The cyber event counts were binned by week, over a time period of $t = 366$ weeks.

This dataset was obtained from a large DoD CSSP that manages computer and network security for multiple defense agencies and organizations working with the DoD (e.g., industry). The cyber events are occurrences of malware that infected the CSSP customer or clients, not against the CSSP itself. Note that this CSSP rarely relies (<1% of cyber event reports) *only* on automated systems for detecting and documenting potential threats. Human analysts examine the data generated by automated systems and collect sufficient evidence to verify true positives; alert data alone are insufficient without human reasoning or validation. While attack data from automated systems generates orders of magnitude more data than analysts, the "attacks" detected by automated systems include false alarms and typically lack human verification and context. Compared to most previous work on cyber attacks, we likely have higher data quality (fewer false positives) but much lower data quantity. For reasons previously mentioned, the quantity of missed malware is unknown.

In addition, the specific activities performed by CSSPs are highly varied [27,28], so the current dataset may not generalize to all CSSPs. For example, CSSPs can differ in their reliance on automated systems rather than analyst verification. Despite such differences, the majority of CSSPs are responsible for reporting of computer/network security incidents [28]. The generic tasks and workflow for cyber analysts in DoD CSSPs are described in [28] and also [29].

Because the data were from an operational CSSP, we are restricted in providing specific details such as the following:

1) Actual years for the cyber events (data are from after the year 2000)
2) Bins smaller than one week (e.g., events per shift or per day)
3) Specific report contents (e.g., method[s] of detection, evidence for the cyber event [also called testimony], type of malware, and infection vector)
4) Names of the CSSP customers, the number of customers, and separate analysis by CSSP customer (even if unidentified)

Therefore, the primary analysis examined the frequency of cyber events (counts). The raw cyber event report data was binned into counts of the number of reports generated per week. This interval size is large enough to allow the data and results to be publically released while still providing a large number of data points to show variability and still minimize the number of empty intervals (no events in a week). Use of the data was approved by the U.S. Army Research Laboratory's Institutional Review Board. Only the aggregated, de-identified dataset and results were approved for public release by operations security for the U.S. Army Research Lab and by the source of the data, the CSSP.

**Cyber events**

To provide more background and context for analyst detection we define a cyber event and show a hypothetical cyber event report. A cyber event documents a computer security event and can be shared among other network security operations. This is formally defined as the violation of security, acceptable use, or standard security practices [16]. Although there are specific content requirements for the filing of an incident within the U.S. Federal Government, report content for the CSSP included the date of analyst detection and a paragraph-long "testimony section" describing evidence for the cyber event. Other information includes the location or site/customer involved and IP addresses for the target and source of the attack. Figure 1 illustrates a hypothetical example of a cyber event report.

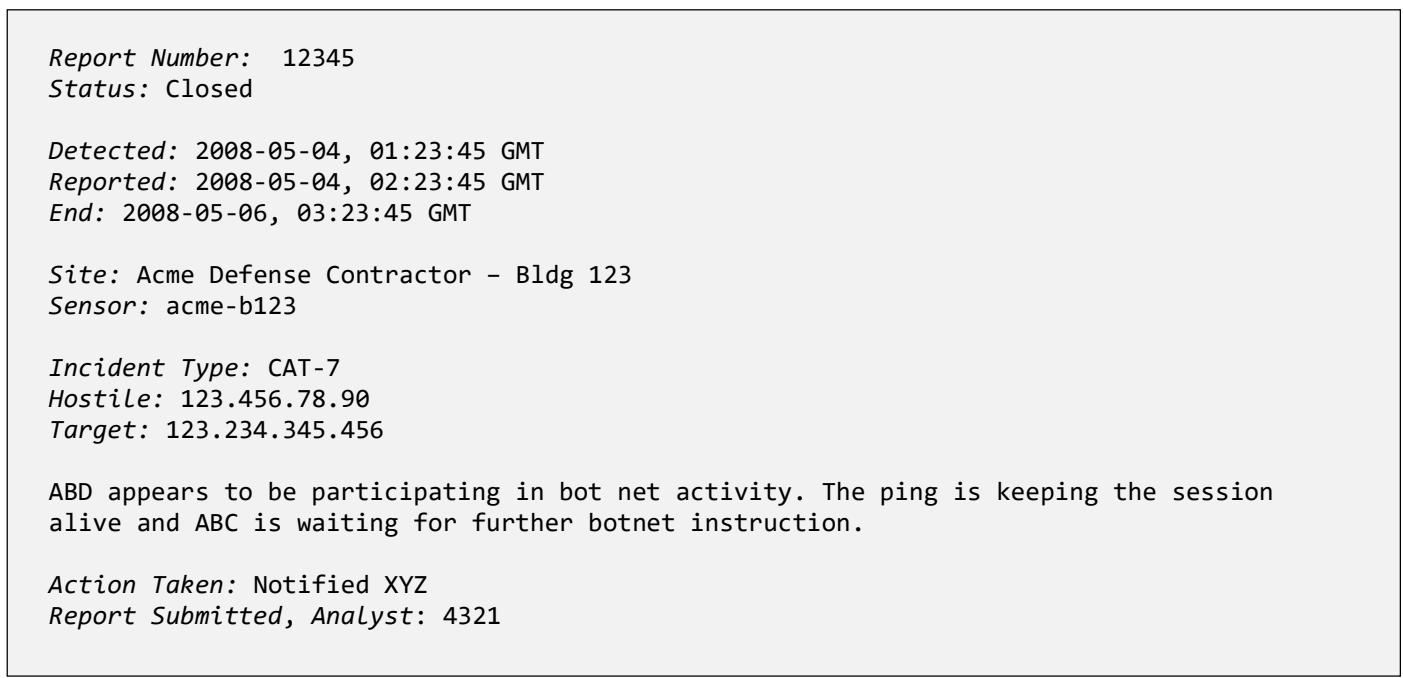

```
Report Number:   12345
Status: Closed

Detected: 2008-05-04, 01:23:45 GMT
Reported: 2008-05-04, 02:23:45 GMT
End: 2008-05-06, 03:23:45 GMT

Site: Acme Defense Contractor – Bldg 123
Sensor: acme-b123

Incident Type: CAT-7
Hostile: 123.456.78.90
Target: 123.234.345.456

ABD appears to be participating in bot net activity. The ping is keeping the session
alive and ABC is waiting for further botnet instruction.

Action Taken: Notified XYZ
Report Submitted, Analyst: 4321
```

**Figure 1.** Hypothetical example of a cyber event report.

An explanation of each of these fields is given to provide greater understanding to the reader.
- Report number – A unique identifier for the incident report
- Status – Indicates if the report is open or closed
- Detected – The initial time that the alert was generated
- Reported – The time that the alert was determined to be significant enough to open a report
- End – Date the issue was resolved
- Site – Location of the attack
- Sensor – Unique identifier of the sensor that captured the alert

- Incident type – Category ranging from 1 to 9, see [30]. CAT-7 (see example above) refers to an incident involving malware or malicious software. Note that we only examine CAT-7 cyber events here.
- Hostile – IP address that issued the command, presumed attacker
- Target – IP address of the host that received the command, defender or potential victim
- Text – Text description of the event(s)
- Action Taken – Describes what action the analyst has taken consistent with CSSP procedures. Typically, the CSSP must notify appropriate POCs at the customer site as well as communicate the findings to peer and superior levels.
- Report Submitted, Analyst – Unique identifier of the analyst who submitted the report

This CSSP has consistent policies and procedures, continuous quality and process monitoring requirements, collaborative reviews, and routine metrics review of how attacks are cataloged. The cyber events that comprise our dataset are all analyst-identified. Each event identified with an analyst report is counted as one instance. Unfortunately, we cannot disclose further details of the counting methodology for security reasons.

## Results

We first summarize characteristics of the dataset: The data were overdispersed and, using a Markov model of burst intensities over time, we found bursts and bursts of bursts. Next, we describe the forecast model. The best-fit forecast model used a BSSM to predict the number of cyber events for a given week using the number from a previous week (one lag model). Taken together, the main results provide compelling evidence that analyst detected cyber events are not point processes (i.e., random values over time). A secondary result was that the annual increase in cyber events was strongly associated with the rising number of CSSP customers each year.

Results are partially reproducible; the data on counts of weekly cyber events, code to reproduce the figures and models, and the Supplementary Materials are available here: https://osf.io/hjffm/. Unfortunately, for the previously mentioned reasons we are unable to share the raw data on cyber events with content, the code used to clean the raw data, and the data used in the secondary result (annual number of CSSP customers).

### Dataset Characteristics

We describe characteristics of the data to summarize it and illustrate the challenges with modeling it. The counts of events were clearly non-normal with overdispersion (Figure 2).

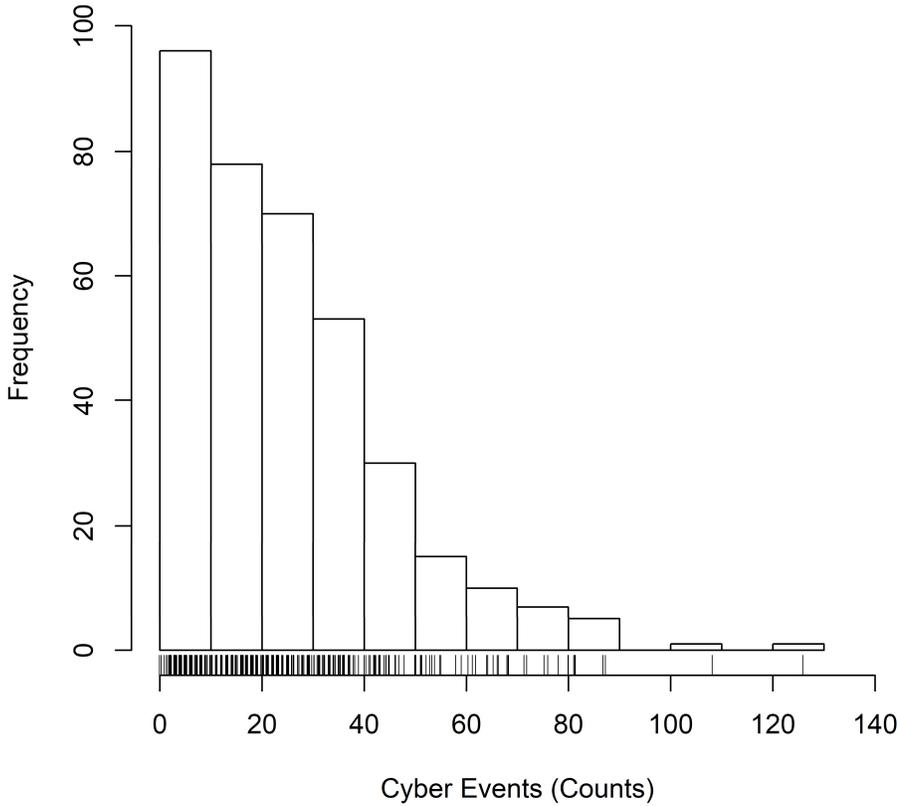

**Figure 2** Figure 2 shows the histogram of cyber events by week. The x-axis is bins of cyber event counts, and the y-axis depicts the number of weeks in each bin. For example, there were 0-10 cyber events in about 90 of the 366 weeks. The tick marks on the x-axis depict the density of cyber events in each bin and their specific values within the bin; these were randomly jittered by 0.3 to create unique values.

To quantify the overdispersion shown in Figure 2, we use the index of dispersion ($I$), also called the variance to mean ratio, shown in Equation 1 [e.g., 20]:

$$I = \frac{\sigma^2}{\mu} \qquad (1)$$

A dispersion index above 1 occurs in data that are overdispersed, because the variance ($\sigma^2$) is greater than the mean ($\mu$) or rate. The index of dispersion in the current data was $I = 15.29$, confirming considerable overdispersion. Many time-series models for count data assume a Poisson distribution (equal rate and variance, expressed in a single parameter), and thus dispersion violations may produce incorrect results [see 23].

*Bursts*

The presence of bursts is strongly suggested by the overdispersion in the data (Figure 2 and the index of dispersion). To detect bursts, we leverage the work of Kleinberg [32] using the rates of cyber events (e.g., 1 cyber event per week has a rate of 1, 10 cyber events per week has a rate of 1/10). Using the Kleinberg model, we find bursts in the rates of cyber events starting at the beginning of Year 3 and also detect bursts of bursts (Level 2 to Level 3), shown in Figure 3. Kleinberg bursts use a Markov model that can simultaneously characterize both the normal and anomalous arrival times: multiple states where each state controls the rate of activities. For our dataset, a higher activity state will exhibit lower rates with shorter amounts of time between attacks (more cyber events), whereas a lower activity state will correspond to increased rates because of longer amounts of time between attacks (fewer cyber events). The Kleinberg model switches between the states with a fixed probability that is independent of the state transitions for previous attack rates. For example, bursts are only detected if there are substantial state transitions from higher activity states to lower activity states and vice-versa. The Kleinberg model generalizes this phenomenon by considering all possible rates as an infinite-state Markov model whose parameters can be learnt from the data rather than by using simple thresholds for detecting bursts. Simpler approaches such as models with a limited number of states and thresholding are generally insufficient for burst detection.

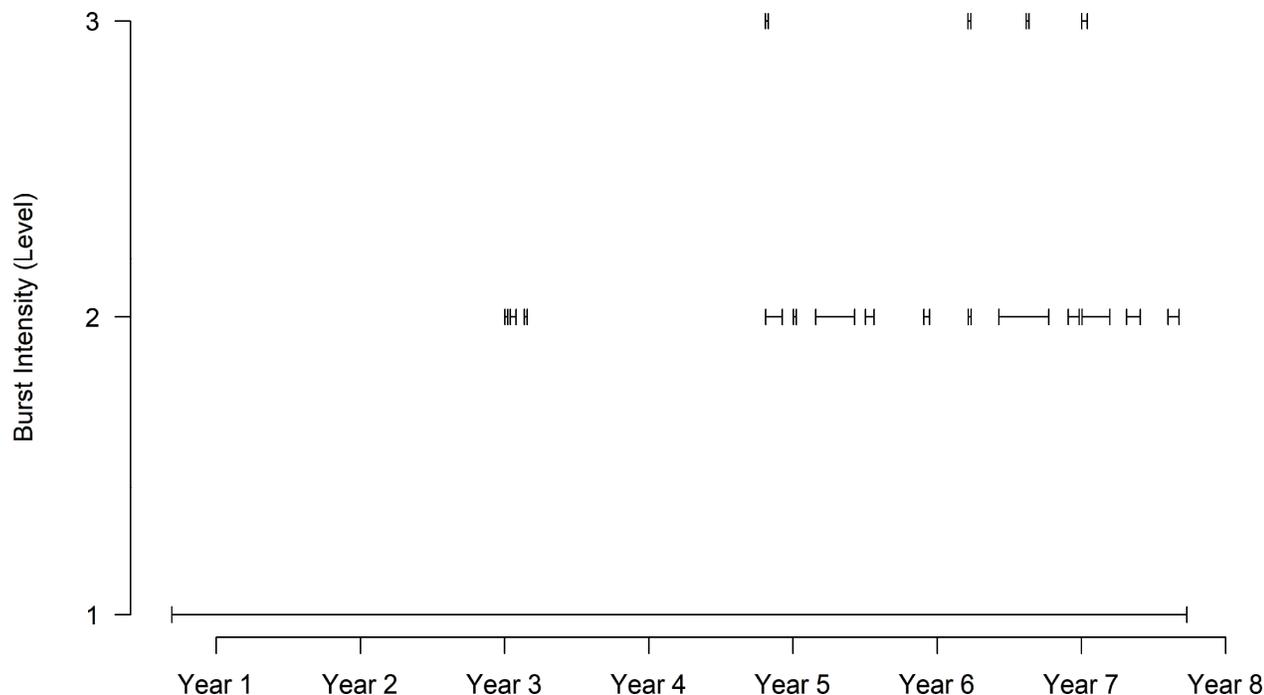

**Figure 3.** The x-axis is the year. The y-axis is the burst intensity level, where level 1 is no bursts and higher levels depict more-intense bursts. The bars represent the range of burst rates for a given burst intensity level. Note the nesting of bursts with faster and faster rates of time to cyber events for levels 2 to 3 (increasing burstiness in the numbers of cyber events). There were a total of 8 weeks with Level 3 bursts and 28 weeks with Level 2 bursts (36 out of 366 weeks exhibited Kleinberg bursts)

Recall our aforementioned point that most cyber attacks exhibit both stochastic and deterministic patterns. If the bursts have a deterministic component, then explicitly modeling them is likely to improve forecast accuracy as others have demonstrated [6,9,10]. It is important to understand the nature of bursts in the data, as effects of bursts or extreme values may be underestimated in model fits used in forecasting. Because reporting of cyber events was consistent, it is unlikely that the burstiness was an artifact of how events were documented.

**Forecasting**

Because of the overdispersion and bursts in the data, the Bayesian State Space Model (BSSM) had better predictive accuracy than the alternative, traditional approaches to time-series forecasting (see Supplementary Materials). The BSSM had several additional advantages over the other approaches to typical time-series forecasting with regression, such as an autoregressive (AR) model, autoregressive moving average (ARMA) model, or autoregressive integrated moving average (ARIMA) model.. First, the BSSM accommodated overdispersed data using a negative binomial distribution rather than assuming a normal distribution. Second, it can estimate multiple sources of variability (e.g., measurement error) [33] instead of only random model variance [34]. Third, BSSM does not require modeling structural breaks (changes in the forecast at specific time points) and detrending data. Fourth, BSSM can accommodate data with non-stationarity (i.e., changes in mean, variance, and correlation structure over time). State space models have become increasingly popular for forecasting problems because of increased computing power and a family of well-developed Markov Chain Monte Carlo (MCMC) algorithms [35]. At a high level, this approach decomposes the underlying generative process into two types of variables:

   a) Observation variables

   b) State variables.

The model then defines a state transition equation that controls how the process moves between states and an observation equation that generates a noisy output based on the current state.

Forecasting was performed using a BSSM with a negative binomial distribution with a one week lag (the same lag as some of the traditional time-series forecasting methods, described in detail in the Supplementary data). BSSM is a transparent machine-learning technique that decomposes the data into observations and the model into simultaneously estimated states [33] (see Equations 2 and 3 and Table 1). The BSSM was implemented in the statistical programming language *R* using the *brms package* [36] as a wrapper for the probabilistic programming language *Stan* [37]. Default, non-informative Bayesian priors, also called Empirical Bayes, were used for model estimation. That is, the priors were empirically estimated from the data. Equation 2 is the overall forecast model for observations and Equation 3 is the state; model variables are defined in the Table I (the equations and variables are from p. 288 in [38]; also see the specification in [33]).

$$\text{Forecast (observation): } y_t = A_t x_t + v_t, \text{ where } y_t \sim NB(n, p) \quad (2)$$

$$\text{Local level (state): } x_t = \phi x_{t-1} + w \quad (3)$$

**Table I: Definitions of Model Variables**

| Variable | Definition |
|---|---|
| $y$ | Estimated number of event(s) |
| $t$ | Time (week) |
| $NB(n, p)$ | $NB$ = negative binomial distribution<br>$n$ = number of failures (zero counts for cyber events)<br>$p$ = probability of success (cyber event occurs) |
| $\phi$ | Latent state |
| $x_t$ | State (transformed known observations from the previous week plus measurement error); note that $x_{t-1}$ is the *known* observation from the previous week |
| $A$ | Measurement/observation matrix |
| $v$ | Measurement error [a] |
| $w$ | Level error [a] |
| | |

a. Note that the credible intervals, described below, incorporates both measurement and level error

No seasonal or cyclical patterns were visually apparent in the data, nor were they meaningful predictors in the alternative models (see the Supplementary data). Consequently, these parameters were omitted from the BSSM model. Including report length of a cyber event as a parameter in the model yielded a worse fit based on the Widely applicable Bayesian Information Criterion (WBIC) [39]. The model with report length had

WBIC = 2590.98, while the model omitting report length had WBIC = 2586.43. Lower WBIC values indicates a better relative model fit.

Figures 4a (weeks for Year 3) and 4b (all 366 weeks) show the one week ahead BSSM forecast. Note that the forecast (white line) in relation to the observed number of events (magenta line). Visible peaks in the magenta line are indicative of bursts in the cyber events. While BSSM leverages deterministic and stochastic patterns, bursts are underestimated in trend lines. Despite deviations in the forecast due to these bursts, nearly all of the observed cyber events were captured by the credible interval (gray-scale shading). However, this assessment of coverage is flawed because we do not know the true values of cyber events which is what credible intervals actually assess. The credible interval (the Bayesian equivalent of a frequentist confidence interval) estimates the probability the true value is captured as a random variable. Formally, the credible interval is defined as the predictive or posterior probability distribution of the model parameters given the observed data [40].

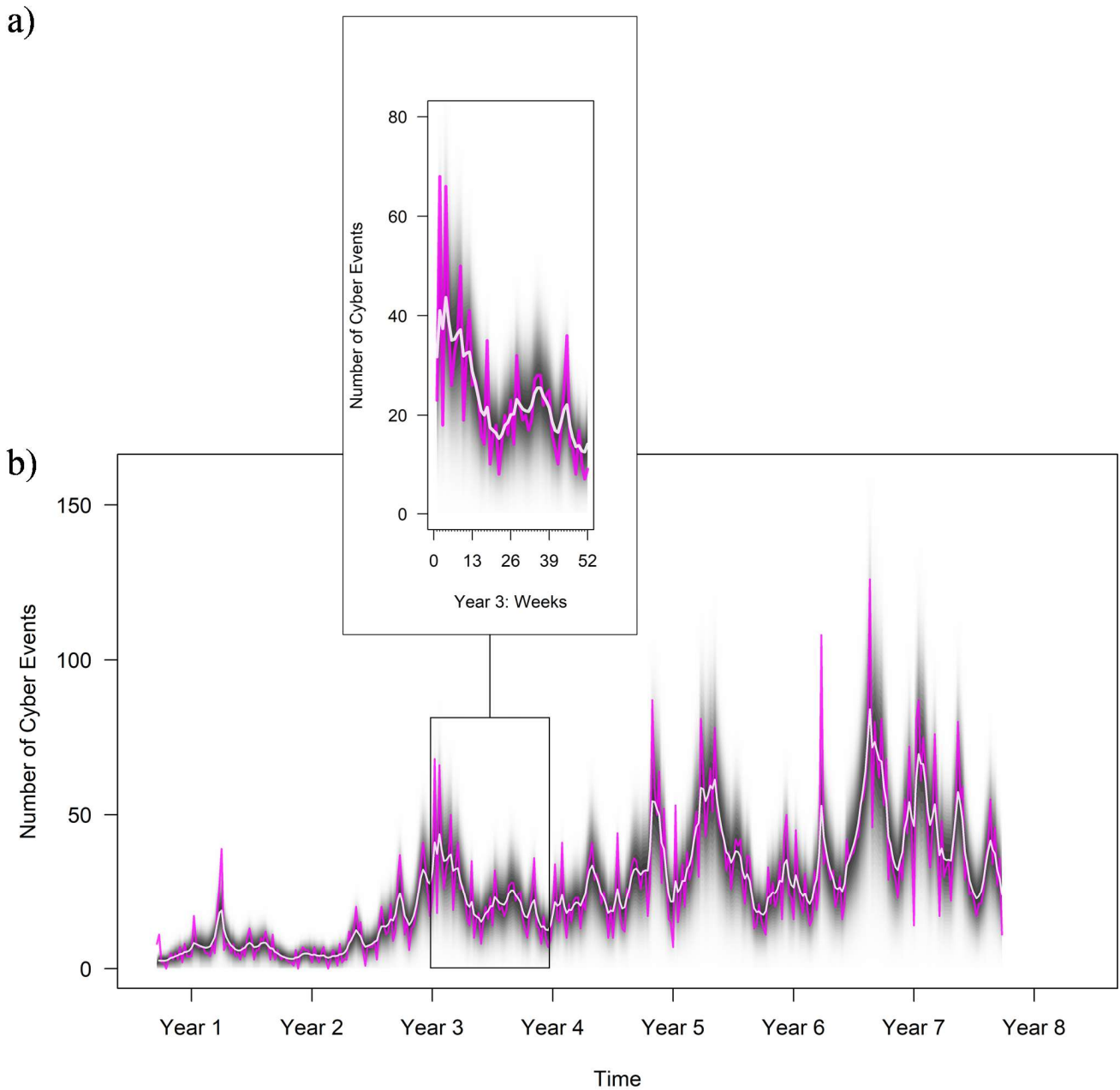

**Figure 4.** a) BSSM forecast for weeks in Year 3; see b) for details. b) BSSM forecast for all 366 weeks. The forecast trend is the white line, a one-week ahead prediction, and the observed number of events is the magenta line. For the model, given the data, the 95% credible intervals depict uncertainty using gray-scale shading. Gradients from dark to light indicate growing uncertainty in estimating the true number of cyber events, but a higher probability that the interval captures the true value.

*Forecast Accuracy*

We quantified the trendline forecast accuracy of the BSSM using multiple measures and compared it to two typical time-series models for all data (Table II) and for only bursts at Level 3 (Table III) and Level 2 (Table IV). For all data and every measure of forecast accuracy, the BSSM outperformed the two other models with overall accuracy that subjectively ranged from decent to excellent. The alternative models had similar accuracy to each other for all data as well as both types of bursts. The BSSM also had higher accuracy than alternative models for the Level 3 and Level 2 bursts. Nevertheless, the trendline accuracy for the BSSM was subjectively very poor for Level 3 bursts and also poor for Level 2 bursts, both with the exception of single measure of forecast accuracy. As a reminder, Kleinberg bursts were prevalent in about 10% of the data (36 weeks out of 366 weeks). The fairly small number of bursts is a challenge for accurate forecasting, even using a model with a distribution for overdispersed counts.. The tradeoffs among different measures of forecast accuracy and equations for each are described by Hyndman and Koehler [41]. Note that the measures of forecast accuracy only reflect the trendline compared to the observed number of cyber events, the true number of cyber events is unknown and can only be estimated

**Table II: Measures of Forecast Trendline Accuracy: All Data**

| Measure | BSSM (one week lag) | AR(1)[a] | AR(3) |
|---|---|---|---|
| Mean Absolute Error (MAE) | 5.43 cyber events | 8.67 cyber events | 8.60 cyber events |
| Mean Absolute Percentage Error (MAPE)[b,c] | 68.17% accuracy | 48.97% accuracy | 49.28% accuracy |
| Symmetric Mean Absolute Percentage Error (SMAPE)[c] | 86.37% accuracy | 80.50% accuracy | 80.50% accuracy |
| Root Mean Square Error (RMSE) | 8.27 cyber events | 13.04 cyber events | 12.98 cyber events |

AR(1) is a first-order autoregressive model: Time $t$ is forecast using $t$-1, a one week lag. AR(3) is a third-order autoregressive model: Time $t$ is forecast using $t$-1, $t$-2, and $t$-3. See Supplementary Materials for details.

a. AR(1) is equivalent to ARMA(1,1)
b. Accuracy is 1 – MAPE and 1 – SMAPE
c. MAPE is undefined with actual values of zero because dividing by zero is undefined. Therefore, we excluded the three weeks with zero cyber events and the fitted trend line corresponding to those three weeks.

**Table III: Measures of Forecast Trendline Accuracy: Level 3 Bursts (highest intensity level)**

| Measure | BSSM (one week lag) | AR(1) | AR(3) |
|---|---|---|---|
| Mean Absolute Error (MAE) | 25.46 cyber events | 37.81 cyber events | 38.05 cyber events |
| Mean Absolute Percentage Error (MAPE) | 42.84% accuracy | 34.75% accuracy | 33.96% accuracy |
| Symmetric Mean Absolute Percentage Error (SMAPE) | 77.84% accuracy | 68.54% accuracy | 68.38% accuracy |
| Root Mean Square Error (RMSE) | 30.53 cyber events | 47.01 cyber events | 47.13 cyber events |

**Table IV: Measures of Forecast Trendline Accuracy: Level 2 Bursts**

| Measure | BSSM (one week lag) | AR(1) | AR(3) |
|---|---|---|---|

| Measure | BSSM (one week lag) | AR(1) | AR(3) |
|---|---|---|---|
| Mean Absolute Percentage Error (MAPE) | 57.01% accuracy | 54.80% accuracy | 55.97% accuracy |
| Symmetric Mean Absolute Percentage Error (SMAPE) | 83.41% accuracy | 81.71% accuracy | 81.75% accuracy |
| Root Mean Square Error (RMSE) | 16.67 cyber events | 21.64 cyber events | 24.61 cyber events |

**Increasing over time: cyber events and CSSP customers**

Figure 4 indicates that the overall number of cyber events is increasing over time. However, we found that this closely corresponds to increases in the number of customers per year. Because the specific information is sensitive, we are unable to provide details. We used the number of events per year as the dependent variable and the number of customers per year as the independent variable in ordinary least squares regression. The regression yielded a strong positive slope, $F(1, 6) = 17.32, p < 0.006$, $Multiple\ R^2 = 0.70$.

## Summary and discussion

We have shown that analyst-detected and -verified cyber events exhibit sufficient systematicity for time-series forecasting, despite overdispersion. We quantified the presence of bursts and their intensity (~10% data). The low prevalence of bursts, across all CSSP customers, appears consistent with findings of bursts for some CSSP customers [12]. Forecast accuracy for bursts was disappointing, even with the BSSM. Nevertheless, the majority of the data could be forecast with reasonable accuracy. The main finding, that the number of cyber attacks from the previous week helps predict the number that will occur one week later, was strikingly similar to other research using human-verified cyber attacks which was from a university network [11]. More broadly, the ability to predict cyber attacks over time is consistent with prior research using processed data from network telescopes and honeypots as well as alerts from automated systems [4,6,7,9–13]. In comparison, we used cyber events detected and verified by analysts in an operational CSSP that rarely used automated systems alone and protects critical infrastructure. Like a weather forecast, albeit a one-week ahead prediction here, cyber event forecasting may proactively enhance threat awareness. This may enable CSSPs and similar organizations to better plan for and manage attacks against their defended domains. Moving from reactive and passive defenses to more proactive defenses may help optimize cybersecurity for both analysts and technical systems.

**Potential applications**

Advance knowledge about the probable range of attack frequency may aid threat awareness in cybersecurity. Potential applications for CSSPs are using the forecast to proactively inform allocation of capabilities such as the sensors and their configuration (e.g., sampling rate, location of sensors) and type of monitoring (e.g., network traffic). Additionally, a cyber event forecast is an empirical estimation of risk which could be directly applied to models of cyber analyst staffing [42,43].

However, we caution that no forecast should be used as a target or a quota for analysts or CSSPs. When a measure becomes a goal, that measure may no longer be meaningful as an outcome [44]. The meaning of a measure can be distorted by biases such as social and political pressure which may introduce incentives with unintended consequences. Reported detection of "more" attacks may not necessarily improve actual computer and network security.

**Limitations**

Because the dataset in this paper is unique, analyst detection and verification of cyber events from an operational CSSP with minimal reliance on automated systems, it also comes with several limitations. First, we have only one dataset from one CSSP. This limits the generalizability of the results, although the central findings were consistent with prior research. Second, even the best fit model had poor accuracy for predicting bursts. This may be because bursts were relatively rare. Third, we treat each cyber event as equivalent. That is, we do not account for differences in impact among attacks (e.g., affected number of computers and/or networks; consequences on security and economic measures such as loss of productivity and time). Attack severity and attack timing may be related. Recent research indicates that for data breaches, there is a meaningful dependency in the timing between attacks and attack severity (magnitude of the data breach) [45]. Fourth, we could not include the report contents (e.g., method[s] of detection, type of malware) in modeling cyber events and were limited to weekly, rather than finer-grained, counts of cyber events. Such information is sensitive because it could reveal how the CSSP monitors and protects systems. While we were unable to do so here, incorporating internal and external variables is likely to improve the quality of the forecast and also aid in identifying factors relevant to specific attacks.

**Factors for attacks?**

Although we can predict malware frequency, we lack direct empirical evidence to explain causes for attack systematicity, as do many others. Moreover, causal inference is a general challenge with observational data [46]. Nonetheless, prediction without identifying causes does not necessarily change the accuracy of prediction. Prediction accuracy will be maintained as long as the conditions and underlying assumptions remain constant. A forecast may become unreliable if conditions change, thus risk models should be frequently recalibrated and validated and should preferably use multiple sources of data [47]. This issue is illustrated by the initial accuracy and then inaccuracy of Google Flu Trends (GFT) [48]. GFT relied on a single source of data (Google search terms related to the flu) and did not update assumptions (e.g., the introduction of suggestions for search terms, other changes to Google search, and media reports) [48].

While we have not identified specific associated or causal factors here, past research suggests that there are multiple causes for cyber attacks. Potential factors are not mutually exclusive. First, there may be planned timing in related attacks (a series of cyber attacks over time) by the same individual or group, or by coordination among groups [49]. In the current work, a series of planned attacks, if they exist, are mixed by aggregation and the absence of detailed information about each event. Recovering separate distributions from their mixtures is challenging [50]. Second, it is possible that exploits are created or purchased on the dark web and deployed by distinct individual or groups around the same time.

Last, activities in social media and events in the physical world likely contribute to attack patterns and vice versa. Prior research has found associations among cyber attacks on DoD networks and foreign media reports of U.S. military actions [51]. Also, website defacements have been linked to a variety of events in the physical world (e.g., violence, protests, and threats) [52]. Incorporating the physical environment as well as expert insights from cybersecurity analysts and analysts in the broader intelligence community may provide additional predictors for cyber attacks and as well as their associations. Given cyberspace is interdependent with the physical environment, adding predictors from experts and other sources could be used to estimate and model interconnected risks among parameters, see [53–55].

**Future directions**

The forecast for analyst detection does not identify specific risk factors associated with attacks. To enhance awareness about specific threats, it is vital to uncover associated and, ideally, causal factors for cyber attacks. This cannot be done with the cyber environment alone because it is inter-dependent with physical environments [1]. In the future, we seek to improve cyber forecasting and to infer the causes for attack patterns. Because of the challenges of openly publishing details with the current dataset, we may use openly available datasets, where attackers sometimes self-identify and even provide the motivation for their attack, such as website defacements (see [52]). Potential variables include events in the physical world as well as more detailed information about the attacks (e.g., the type of malware, exploits/vulnerabilities used,

source[s] of the attack, and malware and other cyber attack pricing on the dark web described in [56]). Also, empirically assessing cybersecurity analysts' understanding of the cyber event forecasts could improve its effectiveness for threat awareness. Research on human understanding of uncertainty in visualization of forecast models is surprisingly limited [57].

Another future direction is combining log and automated defenses and prior knowledge of common vulnerabilities with analyst detection. This could advance understanding how layers of defense are coupled, or not, and how particular attacks pass through layers of defense. Network topology is also relevant to attack forecasting: using log and network topology, an early warning system for mitigating attacks has been developed by modeling probable attack penetration and victims [8].

An additional possibility for future research is using the current forecast models and measures of their accuracy as baselines. We are optimistic that others could develop models that have better accuracy than the BSSM forecast accuracy, especially for bursts. Also, our forecast predictions and assessments of their accuracy were limited to one-week ahead. Future work could evaluate the forecast accuracy of h-step ahead (out of sample) using a variety of models.

A final line of future research is to evaluate alternative loss functions. Most time-series forecasting methods assume the squared loss functions for optimization. However, in the security context, especially for critical infrastructure, investigation of other loss functions is a technical gap. For example, a negative forecast error (i.e., underestimate) could be far more expensive than a positive one (i.e., overestimate). The squared loss function treats both scenarios equivalently. Another line of research is to design forecasting models that can produce richer outputs such as range of forecasts along with the confidence interval or other estimation of uncertainty. Additionally, an interactive forecast model that can output the confidence over a human-specified range is often useful from a risk-aware resource allocation perspective. We also plan to evaluate ensembles of forecasting models so that we can combine the advantages of various forecasting models (e.g., ARIMA, state space-based, and techniques for modeling complex dependencies in the data such as [8]) to produce a superior output.

## Acknowledgments
We thank Bulent Yener, Purush Iyer, and Shouhuai Xu for helpful comments on drafts of the paper, and Mark Gatlin for editing the paper.

The views and conclusions contained in this document are those of the authors and should not be interpreted as representing the official policies, either expressed or implied, of the U.S. Army Research Laboratory or the U.S. government. The U.S. government is authorized to reproduce and distribute reprints for government purposes notwithstanding any copyright notation.

## Funding
This work was supported by the U.S. Army Research Laboratory Postdoctoral Fellowship Program (E.G.Z.) and Senior Fellowship Program (L.R.M.) administered by the Oak Ridge Associated Universities under Cooperative Agreement Number **W911NF-17-2-0003**.

## Conflicts of interest
The authors have no conflicts of interest to declare.

# Supplementary data

In addition to the BSSM model, we also performed traditional approaches to time-series forecasting. Unlike the BSSM model, the alternative models required pre-processing of data to meet the assumption of stationarity (no trending, mean and variance do not change over time, and no structural breaks). Several characteristics of this time series are immediately obvious from visual inspection (see the magenta line in Figure 3b). First, the number of cyber events tends to rise over time. Second, there appears to be a substantial shift in the time series in the early part of Year 2, where both the mean and variability increase (a structural break in the time series). Both trending and structural breaks represent a violation of stationarity. Stationarity is a requirement for most time-series models.

Regression analysis confirmed existence of a linear trend with time, as well as a structural break at time point $t = 83$ (we tested for but did not find evidence of seasonality, another example of non-stationarity). Using the residuals from this regression produces a detrended, stationary time series with mean of zero. We used this detrended time series in the subsequent analyses.

After detrending, we tested for the presence of serial correlation in the data. Inspection of the autocorrelation function (ACF) indicated the presence of strong positive serial correlation. The numbers of cyber event reports in one week tend to be followed by similar numbers the next week.

We also examined number of widely used performance measures for evaluating forecast accuracy. The most direct measure of forecast accuracy is the forecast error: computed as the difference between predicted and observed values. The mean absolute error (MAE) is calculated as the average of the absolute values of forecast errors. The mean squared error (MSE) is calculated as the average of the squared forecast errors. This measure often penalizes large errors. The root mean squared error, or RMSE, is computed as the square root of MSE. Additionally, in the context of our application, predictive accuracy is critical: Observed values appropriately being captured by estimates of interval uncertainty in the forecast models.

For the cyber event prediction problem, we evaluated a number of standard forecasting techniques. However, due to the inherent complexity of the cyber event data, many of them provided poor results. The simplest approach is the moving average (MA) that has a single parameter $k$. The forecast of the model is simply the average of the last k observed values. The weighted moving average method (WMA) assigns different weights to the last $k$ values, with recent values often getting higher weights. The simple exponential smoothing (ES) method predicts the next value as a weighted average of last observed value and the last predicted value with the weights often chosen to minimize the forecast error. The Holt-Winter method (H-W) extends ES to also include seasonality by combining forecasting with three smoothing equations for level, trend, and seasonality respectively. However, we observed that none of these techniques provided appropriate accuracy.

Using ARIMA (autoregressive integrated moving average) models can account for many forms of serial correlation and generate forecasts of future values. The "integrated" or "I" part of ARIMA controls for unit root processes (another type of non-stationarity) by differencing the data. We found no evidence that our time series was a unit root process, and so we restricted our model selection to ARMA models.

Using both the Box-Jenkins approach involving inspection of the ACF and partial ACF [2], as well as automated ARMA selection tools [13], we converged upon an ARMA model of order (1,1). Testing the residuals from this model indicate that it accounts for all the serial correlation in the data, although the residuals do exhibit strong evidence of non-normality (a common feature of count data).

While ARIMA models provided acceptable overall accuracy, they underestimated spikes in events, and the lower bound of the confidence intervals was less than zero. Given the stringent requirements in an operational CSSP, we sought a more accurate forecast using a state space model that captured the distribution of the data. Although ARIMA models can be written as state space models, rather than using ordinary least squares (OLS) regression, the opposite does not necessarily hold.